\documentclass[]{raa}

\usepackage{graphicx,times}
\usepackage{natbib}
\usepackage{amssymb,amsmath}
\bibpunct{(}{)}{;}{a}{}{,}

\usepackage[pagebackref=true]{hyperref}

\usepackage{xspace}

\begin{document}

   \title{TOI-1408: the limb darkening models, photometric and Doppler noises, and the exoplanetary system}

 \volnopage{ {\bf 20XX} Vol.\ {\bf X} No. {\bf XX}, 000--000}
   \setcounter{page}{1}

   \author{Roman V. Baluev\inst{1,2}}

   \institute{Saint Petersburg State University, 7--9 Universitetskaya nab., St Petersburg 199034, Russia;
    {\it r.baluev@spbu.ru}\\
        \and
              Special Astrophysical Observatory, Russian Academy of Sciences, Nizhnii Arkhyz, 369167, Russia\\
\vs \no
   {\small Received 2026 May 19; accepted 2026 Oct 1}
}

\abstract{We perform a self-consistent analysis of Doppler and photometric TESS and ground-based transit observations available
for TOI-1408, known to host a system with two planets orbiting near 2:1 mean-motion commensurability and a distant companion. In
general we confirm its orbital configuration derived in previous studies, however we found that the likelihood function contains
multiple local maxima spread over a wider parametric domain, thus making some model uncertainties significantly larger than
deemed before. In particular, the planet~\emph{b} radius still appears poorly constrained (from $\sim 1.2$~$R_{\rm Jup}$ to $\sim
3.5$~$R_{\rm Jup}$) because of the near-tangential transit geometry. We also show that the derived model is very prior-dependent
regarding the astrophysical (spectrum-derived) constraints on the stellar mass $M_\star$ and radius $R_\star$. While such
estimations may potentially contribute a significant information to the fit, their prior distribution must be accurately
specified to be useful. Presently so accurate characterization (e.g. the correlation between $M_\star$ and $R_\star$) is not
available in the literature.
\keywords{techniques: photometric --- techniques: radial velocities --- methods: data analysis --- planetary systems --- stars: individual: TOI-1408
}
}

   \authorrunning{R.~V. Baluev et al. }            
   \titlerunning{On the TOI-1408 exoplanetary system}  
   \maketitle

%
\section{Introduction}           
\label{sect:intro}

TOI-1408~\emph{b} is a TESS-detected planet\footnote{\url{https://exofop.ipac.caltech.edu/tess/view_toi.php}} with an orbital
period of $P_b\simeq 4.42$~d and a radius initially estimated by $r_b = 0.700\pm 0.055$~$R_{\rm Jup}$. It was confirmed by
\citet{Galazutdinov23} based on Doppler measurements acquired in 2021-2023 by the FFOREST spectrograph mounted at the SAO RAS 6-m
telescope BTA-6 \citep{Valyavin22a,Valyavin22b}. According to them, the planet mass appeared to be $1.69\pm 0.20$~$M_{\rm Jup}$,
but the planet radius was ill-fitted because of the tangential transit geometry, with a low limit of about $1$~$R_{\rm Jup}$
(thus the initial TESS-derived $r_b$ was rejected). Additionally, the Doppler curve implied a significant non-sinusoidal shape
which was consistent with the orbital eccentricity of $e_b\simeq 0.25$. It was also guessed that such an eccentricity could
possibly indicate a presence of other unseen companions in the system. This analysis did not consider potential transit timing
variations (TTVs).

Later on, \citet{Korth24} published a massive set of new Doppler observations and ground-based photometry for TOI-1408 obtained
at multiple sites, as well as their analysis. They argued that there are two planets in the 2:1 mean motion resonance, with the
second small planet~\emph{c} with $r_c = 2.22\pm 0.06$~$R_\oplus$ moving on an internal orbit. The second planet is transiting as
well, although its transits cannot be detected individually. The planets are gravitationally perturbing each other, causing
observationally detectable TTVs with the amplitude of $\sim 8$~min on planet~\emph{b} and $\sim 6-8$~hrs on planet~\emph{c}. 
Their results in fact confirm the initial guess by \citet{Galazutdinov23} about the existence of additional planets. However,
contrary to them \citet{Korth24} obtained a robustly determinable planet radius corresponding to $r_b=2.2\pm 0.4$~$R_{\rm Jup}$.
Additionally, \citet{Korth24} noticed a significant nonlinear Doppler trend indicating the presence of an outer long-period body,
however its orbital parameters and mass could not be determined reliably.

As we can see, the treatment of this system appears rather complicated and involving multiple subtle factors that may potentially
affect the results significantly. The following issues have so far been insufficiently studied.
\begin{enumerate}
\item The limb darkening law should adequately reproduce the near-limb range, because of the near-tangential geometry of the
planet~\emph{b} transit. Notice that \citet{Korth24} tried to fit the limb darkening coefficients directly from the photometric
observations, but this resulted in ill-fitted model in fact. The limb darkening is important to know to correctly assess
exoplanetary radii.
\item The star's astrophysical (spectrum-derived) mass and radius assessments appear quite different in \citet{Galazutdinov23}
and \citet{Korth24}. Probably both are not entirely reliable. It is necessary to verify these estimations and investigate how
this information actually affects the derived system parameters.
\item Doppler and photometric noises, possible systematic errors in the data were not investigated in previous works in
sufficient detail. While \citet{Korth24} tried to reduce stellar rotation by adding a sinusoidal stellar rotational signal, the
actual quality of such a model (e.g. the residual noises, including non-white noise) was not assessed in sufficient detail.
\item Additional subtle but potentially important effects need to be assessed, for example the R\o{}mer effect (variable light
arrival delay) that the outermost body induces on the inner two-planet subsystem.
\end{enumerate}

The goal of the present work is to perform a more detailed analysis treating the issues listed above, and also taking into
account new TESS observations of TOI-1408 that appeared since \citet{Korth24}. The paper structure is as follows. In
Sect.~\ref{sec:ldark} we investigate the limb darkening models for TOI-1408. Sect.~\ref{sect:data} describes the input Doppler
and photometric data and their preliminary analysis, in particular their noises. Sect.~\ref{sect:system} presents the
self-consistent analysis of these data as well as the derived orbital configuration of the planetary system.

\section{The limb darkening law for TOI-1408}
\label{sec:ldark}
The limb darkening law (LDL hereafter) is a mandatory element in any model of exoplanetary transit. There are several standard
functional forms of LDLs, for which numerically computed coefficients are available, tabulated on a grid of stellar spectral
parameters. For example, one may consider a pair of LDL formulae from \citet{ClaretBloemen11,Claret17}:
\begin{equation}
\frac{I_2(r)}{I(0)} = 1 - a (1-\mu) - b (1-\mu)^2, \qquad
\frac{I_5(r)}{I(0)} = 1 - \sum_{k=1}^4 a_k \left(1-\mu^{\frac{k}{2}}\right),
\label{LD}
\end{equation}
where $\mu=\sqrt{1-r^2}$ and $r$ is the distance from stellar disc center (assuming $R_{\rm star}=1$). The first LDL is the
well-known quadratic model, while the other one will be hereafter called as `quartic' for shortness. In the context of planetary
transits we should normalize $I_k(r)$ by the total flux $F_k = 2\pi \int\limits_0^1 I_k(r) r dr$ rather than by $I(0)$, so let us
define renormalized intensity functions $L_k(r) = \pi I_k(r)/F_k$:
\begin{equation}
L_2(r) = \frac{1 - a (1-\mu) - b (1-\mu)^2}{1 - \frac{a}{3} - \frac{b}{6}}, \qquad
L_5(r) = \frac{1 - \sum_{k=1}^4 a_k \left(1-\mu^{\frac{k}{2}}\right)}{1 - \frac{a_1}{5} - \frac{a_2}{3} - \frac{3}{7} a_3 - \frac{a_4}{2}}.
\label{LDnorm}
\end{equation}

Such theoretically constructed LDLs can be helpful but may involve systematic errors. For example, \citet{Baluev15a} revealed
minor systematic biases in the $a$ and $b$ values from \citet{ClaretBloemen11}, as compared to the ones obtained by fitting
transit data directly. However, the latter approach of direct fitting does not always work well. In the case of TOI-1408 an
attempt to fit the quadratic LDL without resorting to theoretic predictions for $a$ and $b$ was made by \citet{Korth24}, but they
obtained an ill-fitted value for $b$. Unfortunately, possible errors in theoretical $I_k$ or $L_k$ models are difficult to
predict because of their complicated nature. One can list the following three major levels where such errors can appear:
\begin{enumerate}
\item uncertainties of the star spectral parameters $T_{\rm eff}$, $\log g$, [Fe/H], and $v_{\rm turb}$;
\item inaccuracy of the analytical law~(\ref{LD},\ref{LDnorm}) approximating numerically computed LDL;
\item inaccuracies in the stellar atmosphere physics assumed when computing that raw LDL.
\end{enumerate}
Here we aim to characterize all three types of the LDL errors in the TOI-1408 case; for shortness they are now referred to as
`level-1', `level-2', and `level-3' errors.

Possible level-1 errors can be characterized by Monte Carlo simulations based on an estimation of spectral parameters ($T_{\rm
eff}$, $\log g$, [Fe/H], $v_{\rm turb}$). Here we used a remodelling of HARPSN spectra from a forthcoming work
\citet{Galazutdinov26} as $T_{\rm eff}=6340\pm 40$~K, $\log g=4.22\pm 0.10$, $\mathrm{[Fe/H]}=0.26\pm 0.1$, and $v_{\rm
turb}=1.4\pm 0.1$~km/s.\footnote{\citet{Korth24} gave some lower $T_{\rm eff}$ and $\log g$, and implicitly assumed $v_{\rm
turb}=1.19$~km/s relying on a calibration from \citet{Bruntt10} based on just $T_{\rm eff}$ (J.~Korth et al., private comm.)} We
considered $18$ photometric passbands (detailed in Table~\ref{tab:ldc}), and for each of them we generated a random sample
containing $10^4$ of $4$-dim vectors of spectrum parameters. We assumed these parameters to follow Gaussian distributions, with
the uncertainties given above, and that different parameters are uncorrelated.\footnote{The correlations are not available in the
literature, though they probably are non-zero, as well as the distributions are probably non-Gaussian. However, such assumptions
are admissible here as we aimed to assess just the order of magnitude of the resulting errors.} For each simulated vector we
interpolated the tables by \citet{ClaretBloemen11} and \citet{Claret17} and computed the coefficients for the LDLs $L_2$ and
$L_5$. So far we considered only the ATLAS atmospheric models. Notice that for the quadratic LDL two alternative pairs of
coefficients are available in the tables, obtained using the flux conservation and least-square methods (FCM and LSM; see cited
works for details). In total we therefore computed $8$ coefficients per a Monte Carlo trial. The interpolation was performed
based on the JKTLD code by \citet{Southworth15} which was augmented to process newer tables and to also post-interpolate with
respect to [Fe/H] and $v_{\rm turb}$.

After that for each LDL coefficient we computed its sample mean and sample standard deviation (SSD). The latter one characterizes
the required level-1 uncertainty of a coefficient. For the majority of passbands these uncertainties appeared quite small, namely
for $L_2$ they were about $0.01$ or below. For $L_5$ coefficients uncertainties appeared some larger, see Table~\ref{tab:ldc}
(first five columns). But the accuracy of the coefficients is not very indicative of the LDL accuracy itself, so we additionally
computed the maximum relative SSDs for $L_2(r)$ and $L_5(r)$, where the maximization was done in some `practical' range for $r$
(see the table and the discussion below). These level-1 error measures are usually about $\sim 10^{-3}$, see sixth column of
Table~\ref{tab:ldc} (they appeared identical for the $L_2$ and $L_5$ LDLs). One can see that for the majority of passbands they
are well below possible relative uncertainties of exoplanets radii in TOI-1408 (e.g. in
\citet{Korth24}). Therefore, level-1 LDL errors for this target appear negligible.

Regarding the level-2 errors, they depend on the LDL form being adopted. According to \citet{Claret17}, the quartic LDL should
have good accuracy for all tabulated spectral parameters and over the entire stellar disc, but this model is difficult for
practical transit modeling because of complicated mathematics. Quadratic LDL is much more practical in this aspect: analytic
formulae are available for the exoplanetary transit curve along with its parametric derivatives \citep{MandelAgol02,AbubGost13},
as well as for the Rossiter-McLaughlin effect \citep{BaluevShaidulin15}. However, the quadratic LDL may not fit the intensity
over the entire stellar disc equally well. In particular, its error may increase near the limb. For TOI-1408 the near-limb domain
is especially important because its planet \emph{b} has near-tangential transit geometry. Such LDL errors may bias the planet
radius estimation.

\begin{figure}
\centering
\includegraphics[width=\textwidth, angle=0]{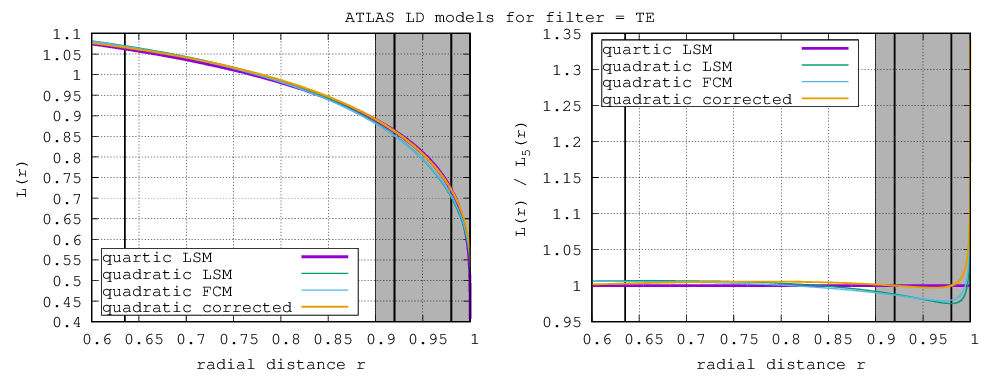}
\caption{Comparing several LDLs for the ATLAS models. See text for a detailed explanation.}
\label{fig:ld}
\end{figure}

Fig.~\ref{fig:ld} demonstrates several LDLs in the range of larger $r$. They correspond to the ATLAS model again and to the TESS
passband (but everything remains visually similar for the majority of other passbands). A vertical line is shown at the abscissa
of the planet~\emph{c} impact parameter $r\simeq 0.635$, which defines the minimum $r$ of practical interest. Also, a near-limb
range $r\geq 0.9$ is highlighted, where the planet \emph{b} transit should reside. As we expected, in the latter range the both
quadratic LDLs (FCM and LSM ones) systematically deviate by $\sim 2$ per cent from the quartic (a more accurate) one. The
deviation is not very big but it may appear important, so we tried to seek for a corrected quadratic LDL that could fit the
desired range better.

For this goal we fixed two reference values of $r$ in this range, namely $r_1=0.92$ and $r_2=0.98$, and solved for such
coefficients $a$ and $b$ that render $L_2(r_{1,2})$ being equal to $L_5(r_{1,2})$. The reference point $r_2$ was chosen where the
spherical atmosphere geometry becomes important (see below), and $r_1$ was chosen symmetrically near another end of the range.

\begin{table}
\bc
\begin{minipage}[]{100mm}
\caption[]{Quartic and corrected quadratic LDLs for TOI-1408 in $18$ passbands\label{tab:ldc}}\end{minipage}
\setlength{\tabcolsep}{1pt}
\small
\begin{tabular}{|c|r@{}lr@{}lr@{}lr@{}l|c|r@{}lr@{}l|c|}
\hline
 Filter &\multicolumn{2}{c}{$a_1$}&\multicolumn{2}{c}{$a_2$}&\multicolumn{2}{c}{$a_3$}&\multicolumn{2}{c|}{$a_4$}&Lev-1&\multicolumn{2}{c}{$a$}&\multicolumn{2}{c|}{$b$}&Lev-2\\
        &\multicolumn{8}{c|}{(uncertainties are level-1, given in brackets)}   & error$^{*}$ & \multicolumn{4}{c|}{(uncertainties are lev-1)} & error$^{*}$ \\
\hline
TESS & $ 0.516$ & $(33)$ & $ 0.268$ & $(85)$ & $-0.137$ & $(104)$ & $ 0.000$ & $(43)$ & $0.0048$ & $ 0.209$ & $(8)$ & $ 0.312$ & $(4)$ & $0.0057$ \\
\multicolumn{15}{|c|}{Sloan}\\
   u & $ 0.348$ & $(55)$ & $ 0.519$ & $(190)$ & $ 0.209$ & $(232)$ & $-0.198$ & $(85)$ & $0.0135$ & $ 0.566$ & $(28)$ & $ 0.239$ & $(23)$ & $0.0029$ \\
   g & $ 0.350$ & $(28)$ & $ 0.684$ & $(98)$ & $-0.227$ & $(133)$ & $ 0.013$ & $(55)$ & $0.0089$ & $ 0.458$ & $(14)$ & $ 0.274$ & $(9)$ & $0.0050$ \\
   r & $ 0.448$ & $(30)$ & $ 0.546$ & $(86)$ & $-0.329$ & $(115)$ & $ 0.062$ & $(48)$ & $0.0066$ & $ 0.284$ & $(10)$ & $ 0.327$ & $(5)$ & $0.0061$ \\
   i & $ 0.515$ & $(33)$ & $ 0.284$ & $(86)$ & $-0.152$ & $(107)$ & $ 0.005$ & $(44)$ & $0.0051$ & $ 0.209$ & $(8)$ & $ 0.315$ & $(4)$ & $0.0058$ \\
   z & $ 0.560$ & $(34)$ & $ 0.080$ & $(81)$ & $ 0.004$ & $(95)$ & $-0.049$ & $(38)$ & $0.0039$ & $ 0.160$ & $(7)$ & $ 0.302$ & $(4)$ & $0.0054$ \\
\multicolumn{15}{|c|}{Johnson-Cousins}\\
   U & $ 0.368$ & $(55)$ & $ 0.431$ & $(185)$ & $ 0.313$ & $(223)$ & $-0.232$ & $(81)$ & $0.0122$ & $ 0.582$ & $(27)$ & $ 0.223$ & $(22)$ & $0.0025$ \\
   B & $ 0.348$ & $(32)$ & $ 0.574$ & $(112)$ & $ 0.004$ & $(144)$ & $-0.078$ & $(57)$ & $0.0092$ & $ 0.534$ & $(16)$ & $ 0.233$ & $(11)$ & $0.0039$ \\
   V & $ 0.403$ & $(29)$ & $ 0.642$ & $(91)$ & $-0.339$ & $(122)$ & $ 0.064$ & $(51)$ & $0.0074$ & $ 0.354$ & $(12)$ & $ 0.311$ & $(6)$ & $0.0059$ \\
   R & $ 0.463$ & $(31)$ & $ 0.488$ & $(86)$ & $-0.296$ & $(112)$ & $ 0.052$ & $(47)$ & $0.0062$ & $ 0.262$ & $(10)$ & $ 0.326$ & $(4)$ & $0.0061$ \\
   I & $ 0.535$ & $(34)$ & $ 0.206$ & $(86)$ & $-0.090$ & $(104)$ & $-0.018$ & $(42)$ & $0.0047$ & $ 0.192$ & $(8)$ & $ 0.312$ & $(4)$ & $0.0056$ \\
   J & $ 0.509$ & $(24)$ & $ 0.296$ & $(56)$ & $-0.457$ & $(67)$ & $ 0.174$ & $(29)$ & $0.0032$ & $ 0.066$ & $(4)$ & $ 0.316$ & $(2)$ & $0.0067$ \\
   H & $ 0.730$ & $(9)$ & $-0.195$ & $(19)$ & $-0.161$ & $(19)$ & $ 0.113$ & $(8)$ & $0.0026$ & $-0.006$ & $(2)$ & $ 0.310$ & $(3)$ & $0.0073$ \\
   K & $ 0.556$ & $(16)$ & $-0.025$ & $(39)$ & $-0.244$ & $(49)$ & $ 0.125$ & $(21)$ & $0.0023$ & $ 0.003$ & $(3)$ & $ 0.264$ & $(1)$ & $0.0060$ \\
\multicolumn{15}{|c|}{Str\"{o}mgren}\\
   u & $ 0.313$ & $(52)$ & $ 0.622$ & $(176)$ & $ 0.092$ & $(215)$ & $-0.146$ & $(80)$ & $0.0143$ & $ 0.576$ & $(27)$ & $ 0.236$ & $(21)$ & $0.0031$ \\
v$^{**}$& $ 0.315$ & $(35)$ & $ 0.810$ & $(92)$ & $ 0.185$ & $(165)$ & $-0.155$ & $(64)$ & $0.2313$ & $ 0.895$ & $(58)$ & $ 0.194$ & $(15)$ & $0.0034$ \\
b$^{**}$& $ 0.307$ & $(26)$ & $ 0.532$ & $(130)$ & $-0.340$ & $(130)$ & $ 0.057$ & $(54)$ & $0.0267$ & $ 0.197$ & $(25)$ & $ 0.274$ & $(9)$ & $0.0045$ \\
   y & $ 0.401$ & $(29)$ & $ 0.646$ & $(94)$ & $-0.333$ & $(126)$ & $ 0.060$ & $(53)$ & $0.0076$ & $ 0.361$ & $(12)$ & $ 0.309$ & $(6)$ & $0.0058$ \\
\hline
\end{tabular}\\
\ec
\tablecomments{0.86\textwidth}{
$^{\phantom{*}*}$ See text for the definition of error measures.
$^{**}$ These lines may be unreliable; see text for a discussion.
}
\end{table}

Although this is rather simplistic approach, the corrected LDL behaves considerably better in the desired domain. Notice that so
far we ignored the range $r\geq r_2$. It is affected by the effect of stellar surface spherical curvature, while all ATLAS-based
models neglect this effect and are expectedly inaccurate there.

Simultaneously one should pay some attention to the wider $r$ range covered by only the planet \emph{c} transit, $0.635<r<0.9$.
It is important that the maximum LDL error in this range did not get increased after our correction.

In Table~\ref{tab:ldc}, last three columns, we provide coefficients for this corrected LDL, for all $18$ photometric passbands.
They are accompanied by an additional `level-2' error measure, which represents the maximum of
$\left|\log\frac{L_2(r)}{L_5(r)}\right|$ in the same `practical' $r$-range as above. Notice that level-2 errors for $L_5$ were
undefined because there is no any better model that $L_5$ could be compared to.

For the majority of passbands both LDL error measures are below $1$ per cent, or at least about that. However, in case of the
Str\"omgren $v$ filter the level-1 error got suddenly big while some coefficients look odd, as compared to the neighbouring
filters. Moreover, the quartic LDL for this filter appears negative at $r=1$ (and near it). We verified that our interpolation
was not the cause of this issue, and it comes from the original tables by \citet{ClaretBloemen11}. For the Str\"omgren $b$ and
$v$ bands about a quarter of tabulated lines imply $L_5(1)<0$, although in other bands such occurrences are rare or not present at
all. We are not in position to explain this behaviour, so we just conclude that $b$ and $v$ filters LDLs may be unreliable.

\begin{figure}
\centering
\includegraphics[width=\textwidth, angle=0]{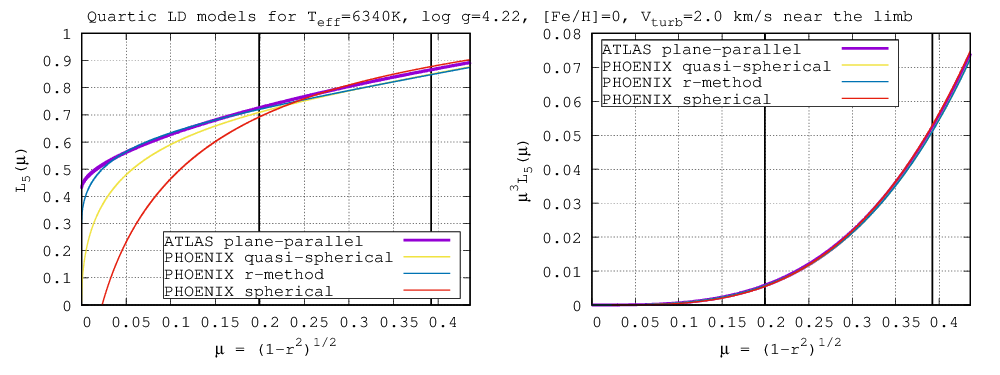}
\caption{Near-limb behaviour of several LDLs assuming different atmosphere models. See text for a detailed
explanation.}
\label{fig:ldl}
\end{figure}

Yet we need to discuss the LDL level-3 error. It can reveal itself by comparing the LDLs derived from different atmospheric
models. For the TESS passband \citet{Claret17} provided a set of tables for the alternative PHOENIX models that take into account
spherical geometry of stellar atmosphere (while the ATLAS tables always assume plane-parallel geometry). These additional tables
only cover solar metallicity and $v_{\rm turb}=2$~km/s, so they do not apply to the majority of stars, including TOI-1408.
However, they are useful to get impression of the magnitude of the effect. We interpolated all PHOENIX and ATLAS tables to our
nominal values of $T_{\rm eff}$, $\log g$ for TOI-1408 but setting the metallicity to be solar one and $v_{\rm turb}=2$~km/s. The
corresponding quartic LDLs are shown in Fig.~\ref{fig:ldl}, left panel, as functions of $\mu$. We can see that different models
start to diverge only for $\mu<0.2$, which corresponds to an extremely thin ring $r>0.98$.

Notice that although planet \emph{b} has a near-tangential transit, it usually obscures a much wider $r$ range than this
threshold. Therefore, the spherical effect may occur only in the very beginning and ending parts of the transit, where only the
$r\geq 0.98$ ring is obscured. But in that case the flux drop itself would be small because the area being obscured by the planet
is small. The LDL errors manifest themselves relatively to the flux drop and therefore have a smaller effect on the transit
lightcurve then it may seem from the literal LDL behaviour. From the classic formulae \citep{MandelAgol02,AbubGost13} it follows
that near the limb an obscured domain area decreases as $\sim \Delta r^{\frac{3}{2}}$, where $\Delta r$ stands for the width of
this domain (in radial direction). In this limit the observed flux drop should be roughly proportional to the same factor, and as
long as $\mu \sim \sqrt{1 - (1-\Delta r)^2} \sim \Delta r^{\frac{1}{2}}$, the intensity function $L(\mu)$ should be scaled down
by $\mu^3$. Such a scaled function would roughly measure the observable flux drop, and it is shown in the right panel of
Fig.~\ref{fig:ldl} (for the same set of PHOENIX and ATLAS models as in the left one). All these curves appeared visually
indistinguishable from each other; in fact the area factor $\mu^3$ effectively suppressed all divergences seen in the left panel.
Therefore, the near-limb effect of spherical atmosphere appears practically negligible in the context of TOI-1408 and of modelling
its exoplanets transit curves.

\section{Observations and their noises}
\label{sect:data}

\subsection{Photometry}
\label{sect:phot}
Regarding the TESS observations of TOI-1408, we used the same 12 Sectors as in \citet{Korth24}, but also added newer observations
for six more Sectors 77, 78, 79, 84, 85, and 86, downloadable from the Mikulski Archive for Space
Telescopes\footnote{\url{https://archive.stsci.edu/}}. Notice that several TESS data products are offered that correspond to
different processing levels of the same observations. \citet{Korth24} used the PDC (Presearch Data Conditioning) data for their
analysis. However based on a comparative investigation detailed in \citet{BaluevSokov25} we prefer the DVT (Data Validation
Timeseries) LC\_INIT data. Most of the time these data differ from the PDC ones by quite a small amount, but they correct for
minor edge-jumps effects likely related to the satellite momentum dumps occurring every few days.

\begin{figure}
\centering
\includegraphics[width=\textwidth, angle=0]{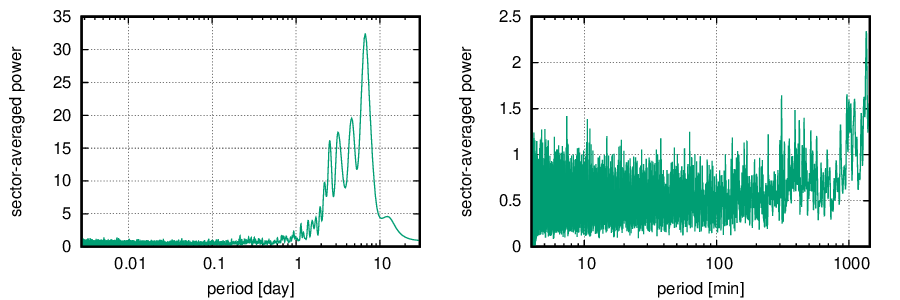}
\caption{Inter-sector averaged periodogram of the TESS out-of-transit data for TOI-1408. Notice that the Nyquist (the minimum
reasonable) period for these data is $4$~min, while the typical duration of a single sector chunk (the maximum reasonable period)
is $\sim 1$~month.}
\label{fig:powTESS}
\end{figure}

Before proceeding to the final analysis, we investigated the quality of these data. For example, we needed to ensure that there
is no significant stellar variability, which was not demonstrated by \citet{Korth24} clearly. Similarly to what was done in
\citep{BaluevSokov25} for WASP-33, we cut away all exoplanetary transits from the TESS data to keep only possible intrinsic
stellar variations. Notice that transit times are not strictly periodic because of mutual planetary perturbations, so we used
transit times derived from a preliminary fit similar to the ones discussed in Sect.~\ref{sect:system} below (we omit its details
as they are not very important). The resulting out-of-transit TESS data were then split per 18 sectors, and for each sector we
computed a DCDFT (or GLS) periodogram \citep{FerrazMello81,ZechKur09}. To keep only the systematic noise we averaged all these
single-sector periodograms between sectors, and such intersector-averaged periodogram is shown in Fig~\ref{fig:powTESS}. We can
see a large power excess in the period range from $\sim 1$~d to $\sim 10$~d, likely indicating some photometric variability (or
possibly instrumental effects). If we analysed TESS data on a per-sector basis we would need to model it in some way. However, we
do not need to model each Sector in full, since we only need to capture individual planetary transits which have a much shorter
duration ($\sim 1.6$~hr for planet~\emph{b} and $\sim 2.6$~hr for planet~\emph{c}). The power excess noticed above is small at
such timescales so it is unlikely to affect planetary transit lightcurves. Therefore, to get rid of this variation we extracted
only limited segments of TESS data centered on each transit ($\pm 2$~hr for planet~\emph{b} and $\pm 3$~hr for planet~\emph{c}).
Inside of each such segment the variability under discussion would manifest itself, at the worst, as a long-term trend which can
be effectively modelled by a polynomial.\footnote{In three cases the planet~\emph{b} and~\emph{c} transits occurred close to each
other so their data segments overlapped. In these cases we merged two segments into a single one containing the both transits;
the merged segments were only slightly longer than the $\pm 3$~hr ones.} We also investigated each of single-sector periodograms
and ensured that none revealed any significant individual peaks inside our sensitive period range.

\begin{figure}
\centering
\includegraphics[width=\textwidth, angle=0]{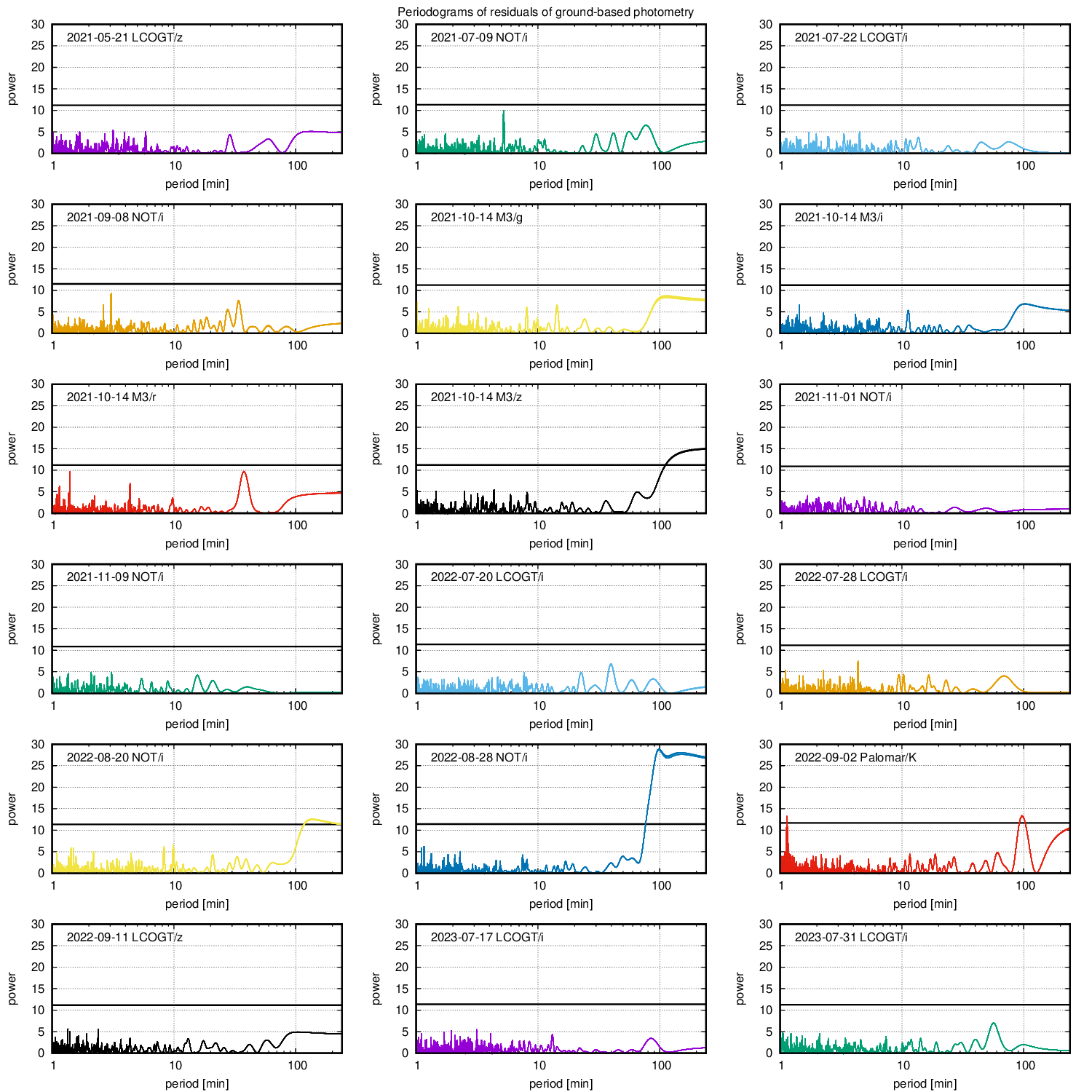}
\caption{Periodograms of residuals of ground-based transit lightcurves of TOI-1408 released by \citep{Korth24}. The telescopes,
filters, and dates of the observations are labelled in each plot. Also, each plot shows a $99\%$ significance level computed
based on \citep{Baluev08a}.}
\label{fig:powextra}
\end{figure}

\citet{Korth24} also released a set of $18$ ground-based lightcurves of planet~\emph{b} transits. We tested these data
for possible presence of variations as well. Based on a preliminary fit we computed the corresponding residuals and performed
their periodogram analysis (see Fig~\ref{fig:powextra}). Four of these lightcurves revealed statistically significant long-term
variations, namely those from the telescopes M3 ($z$-band, taken at 2021 Oct 14), NOT ($i$-band, 2022-08-20 and 2022-08-28),
Palomar ($K$-band, 2022-09-02). Some of these variations may look rather marginal, but nevertheless we decided to remove all the
affected lightcurves from our final analysis. An important notice is that periodograms in Fig.~\ref{fig:powextra} remain
practically independent of the choice of a likelihood maximum discussed in Sect.~\ref{sect:system}.

\subsection{Doppler data}
\label{sect:rv}

First high-precision Doppler data for TOI-1408 were published by \citet{Galazutdinov23}, which contained $19$ RV measurements
acquired by the new FFOREST spectrograph mounted at $6$-m BTA telescope in Special Astrophysical Observatory (SAO) in Russia.
Further on, \citet{Korth24} released a large set of additional Doppler measurements obtained at multiple other sites. In total we
have 12 independent Doppler datasets corresponding to either different telescopes or, in some cases, to different detectors on
the same telescope, or to different spectrum calibration methods. In our self-consistent analysis below we aim to include all
these data.

\citet{Korth24} noticed that Doppler data may be affected by stellar rotation which they modelled by adding ``a sinusoidal
stellar rotational signal''. However, such stellar activity (including rotationally-modulated one) does not usually manifest
itself as a literal sinusoid. Usually it is a quasiperiodic Doppler variation representing an autocorrelated random process. It
is unlikely that modelling such a variation through a deterministic function like sinusoid may offer reliable results. One
standard way to treat such data is the Gaussian Process (GP) method relying on some parametric model for the correlation function
\citep{Baluev11,Baluev13a,Rajpaul15,Foreman-Mackey17,Baluev18c}. However, before that we should investigate the spectrum of this
noise to decide which particular model to choose.

\begin{figure}
\centering
\includegraphics[width=0.99\textwidth, angle=0]{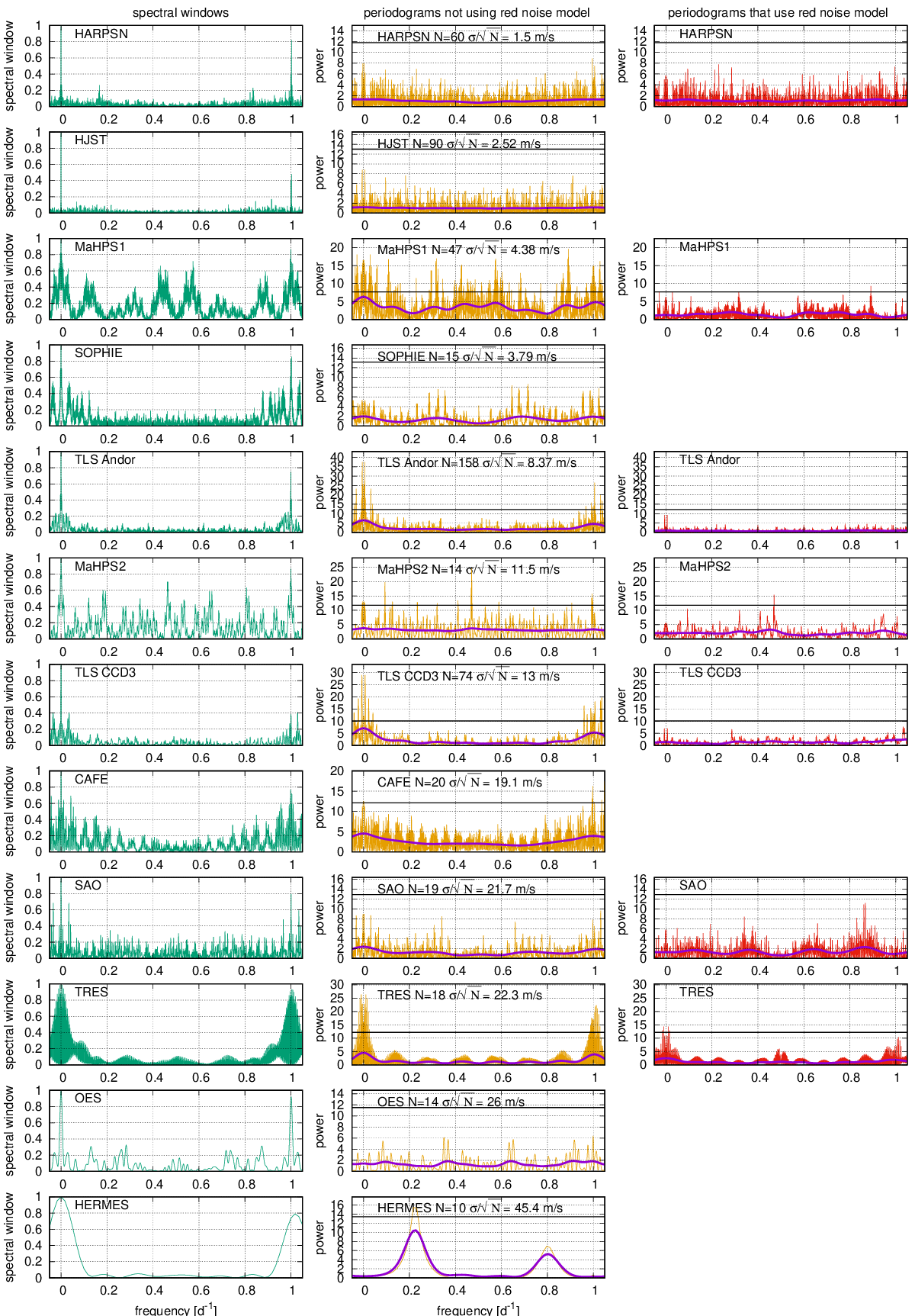}
\caption{Periodograms of residuals for all Doppler datasets of TOI-1408, arranged per-row. Each row shows the spectral window
(left), the periodogram assuming no red noise (middle), and (optionally) the periodogram assuming a red-noise model (right). The
rows are sorted by the parameter $\sigma/\sqrt N$, where $\sigma$ is the residuals r.m.s. and $N$ is the number of observations.
Each periodogram graph also shows a smoothed periodogram and $99\%$ significance level as horizontal line.}
\label{fig:rvpow}
\end{figure}

Based on a preliminary fit that assumed, besides the inner planets \emph{b} and \emph{c}, only white noise in all Doppler data
and a quadratic RV trend to model the long-period companion~\emph{d}, we derived the RV residuals and computed their periodograms
per each of 12 subsets. These are shown in Fig~\ref{fig:rvpow}, middle column. In the most cases there is no significant
individual peak, although many periodograms reveal a clear power excess in different frequency ranges. To verify this we passed
these data through the red noise detection algorithm of {\sc PlanetPack} software \citep{Baluev18c}. A significant red noise was
detected in 7 datasets. However, these periodograms reveal no obvious hint of a consistent stellar rotation period. Moreover, the
power spectrum looks quite differently for different datasets. In the best-accuracy HARPSN dataset the red noise is rather
shallow and its power excess appears widely smeared, thus indicating just a red noise without obvious periodic modulation (which
would look, if present, like an additional power excess centered at a non-zero frequency). This power excess looks reminiscent of
classic examples demonstrating stellar activity like GJ~581 \citep{Baluev13a} or GJ~876 \citep{Baluev11}. However, other datasets
like TLS (``Andor'' and ``CCD3'') or TRES reveal a considerably different power excess with a much larger amplitude and more
narrow frequency localization. Since different datasets reveal clearly different noise spectrum, their underlying physics is
unlikely to be the same. While the correlated noise in HARPSN may indicate some intrinsic stellar activity indeed, we believe
that other datasets with correlated noise may reveal some long-term instrumental effects rather than stellar activity.

Since our periodograms did not reveal clear stellar rotation period we assumed a red noise model with decay-only correlation
function without rotational modulation:
\begin{equation}
k(t_i-t_j) = \sigma_{\rm white}^2(s) \delta_{ij} + \sigma_{\rm red}^2 \exp\left(-\frac{|t_i-t_j|}{\tau_{\rm red}}\right),
\label{ecorr}
\end{equation}
where $t_i$ and $t_j$ are two measurement timings, $\delta_{ij}$ is Kronecker delta, $\sigma_{\rm white}(s)$ and $\sigma_{\rm
red}$ are the white and red noise standard deviations, respectively, and $\tau_{\rm red}$ is the red noise correlation timescale.
The white noise $\sigma_{\rm white}(s)$ depends on the fittable jitter parameter $s$ through a regularized model, see
\citet{Baluev14a}, with the regularization parameter $a$ set to $0.5$.

Different datasets were allowed to have different fittable parameters $s$, $\sigma_{{\rm red}}$, and $\tau_{\rm red}$. The model
shape~(\ref{ecorr}) was kept the same for nearly all datasets, except HARPSN one. In the latter case~(\ref{ecorr}) allowed to
remove the power excess only partially. Since HARPSN data dominate in term of their statistical contribution, we tried harder to
find a neat noise model for it, and after some experimenting we adopted the following correlation function:
\begin{equation}
k(t_i-t_j) = \sigma_{\rm white}^2(s) \delta_{ij} + \frac{\sigma_{\rm red}^2}{1+\frac{(t_i-t_j)^2}{\tau_{\rm red}^2}},
\label{ecorr2}
\end{equation}
which performed better.

Periodograms using such red noise models are shown in the rightmost column of Fig~\ref{fig:rvpow}. Those are likelihood ratio
periodograms as defined by \citet{Baluev13a}. We can see that the majority of power spectra got flattened leaving no
statistically significant peaks. Some formally significant but marginal residual variations still remained in the TRES, two
MaHPS, and CAFE datasets. We believe they can be neglected because these datasets are less accurate and less numerous than e.g.
the HARPSN one, so their statistical contribution is small.

Based on these results, for our final analysis below we adopted generally the same noise models as described above, with some
minor modifications. First, two TLS datasets revealed rather similar red noise with mutually consistent $\sigma_{\rm red}$ and
$\tau_{\rm red}$, however $\tau_{\rm red}$ had very large uncertainty so that only its order of magnitude was reliably
determinable. To avoid possible ill-fit issues in the final analysis we decided to bind $\tau_{\rm red}$ parameters between these
datasets, still keeping $\sigma_{\rm red}$ different. A similar issue appeared in the two MaHPS datasets: the second of them has
very few data and although its red noise appeared significant, its red noise parameters were ill-fitted. So to improve this we
bound the $\tau_{\rm red}$ parameter between these two datasets as well.

\section{Self-consistent data analysis and planetary system parameters}
\label{sect:system}

\subsection{TTV pre-analysis}

\begin{figure}
\centering
\includegraphics[width=\textwidth, angle=0]{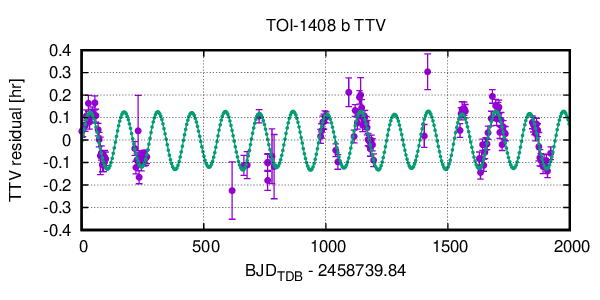}\\
\includegraphics[width=\textwidth, angle=0]{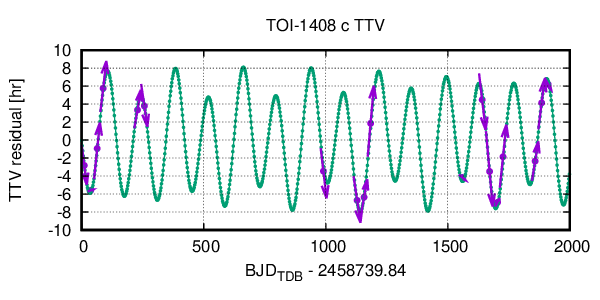}
\caption{TTV signals of TOI1408~\emph{b} (top) and~\emph{c} (bottom) against the TTV model corresponding to one best fit from
Sect.~\ref{sect:fits}. See text for further details on the construction. Uncertainties of vectors in the bottom panel are
comparable to the arrows width and omitted for the sake of clarity.}
\label{fig:ttv}
\end{figure}

The hypothesis of planet TOI-1408~\emph{c} existence was largely based on the interpretation of planet~\emph{b} TTV signal.
Therefore, before proceeding to two-planet models we ought to verify additionally that this TTV interpretation is not affected by
our changes to the noise models or by adding new data. For this goal we considered only lightcurves containing planet~\emph{b}
transits and performed two single-planet fits, namely assuming (i) null (no-TTV) hypothesis with strictly linear model for
planet~\emph{b} transit times and (ii) alternative hypothesis allowing all transit times to be fittable. The null fit appeared
rather similar to the one discussed in \citep{Galazutdinov23}. The second fit (with free transit times) offered an undoubtedly
significant improvement in terms of the likelihood-ratio statistic. Namely, the log-likelihood-ratio was $\sim 450$ corresponding
to the chi-square test statistic of $\sim 900$, which is very large even though the number of degrees of freedom was $108$ (the
same as the number of planet~\emph{b} transit lightcurves). The TTV graph reveals an expected near-sinusoidal variation without
much difference from \citet{Korth24}, see Fig~\ref{fig:ttv} (top panel).

Another intricate issue appears regarding the TTV signal of planet~\emph{c}. This planet's transits are so shallow that they
cannot be reliably detected individually. This was first noticed by \citet{Korth24} who also argued that these transits were only
detectable when analysing a whole-sector data (each TESS sector contains about a dozen of planet~\emph{c} transits). In this case
fitting all such transits independently, as we did for planet~\emph{b}, can be dangerous because we may fit some local noise
fluctuations thinking they are transits. Additionally, planet~\emph{c} demonstrates very large TTV amplitude of $\sim 8$~hr, so
its apparent orbital period varies from sector to sector. Moreover, when analysing the entire data (all sectors combined), the
planet~\emph{c} transit signal gets smeared and becomes undetectable again. To verify this TTV we adopted the following scheme.
First we cut away planet~\emph{b} transits from the TESS data, keeping only planet~\emph{c} ones, and then computed, per each
TESS sector, the box least-squares (BLS) periodogram \citep{Kovacs02} in a narrow period range from $2.15$~d to $2.25$~d (about
the proposed planet~\emph{c} period). For the majority of sectors the BLS periodogram revealed a single dominating maximum, for
which we computed the apparent orbital period $P_s$ along with the best fitting reference transit time $T_s$ (choosing the one
closest to the middle of the sector). Only for Sector~59 the BLS periodogram reveals \emph{two} nearly equally tall maxima.
Notice that in this setting the data passed to the BLS periodogram retain the small stellar variability noticed above (in the
period range $1-10$~d), and it can explain a peculiar behaviour in some sectors.

After that we used the derived per-sector values of $T_s$ and $P_s$ to plot the corresponding TTV as a set of vectors. Each $T_s$
defined the midpoint of a vector, while $P_s$ defined its inclination (the local TTV slope). We found quite a good agreement
between such vectorial TTV graph and the TTV model proposed by \citet{Korth24} as well as with our TTV models (see
Fig~\ref{fig:ttv}). The only clause was regarding the peculiar Sector~59, for which we had to choose the \emph{second} high peak
to bring the corresponding vector close to the TTV model.

Therefore, both TTV signals appear preserved with our noise treatment and both seem to continue the same variations as
were seen in the pre-2024 data.

\subsection{Three-planet model analysis}
\label{sect:3pmod}
Our final three-planet system model involved two inner planets \emph{b} and \emph{c} gravitationally interacting with each other
while the companion~\emph{d} being gravitationally detached, moving on a distant circular orbit. The orbital configuration was
assumed coplanar. Thanks to the gravitational perturbations in the inner subsystem the mass of the star can be estimated purely
dynamically, so in the first treatment attempt we did not use any astrophysical estimations for it, as well as for the star
radius. The limb darkening model, as well as noise models in the photometric and Doppler data, were described above. All the
fitting job was done using {\sc PlanetPack} software \citep{Baluev13c,Baluev18c} implementing the likelihood maximization.

\begin{figure}
\centering
\includegraphics[width=\textwidth, angle=0]{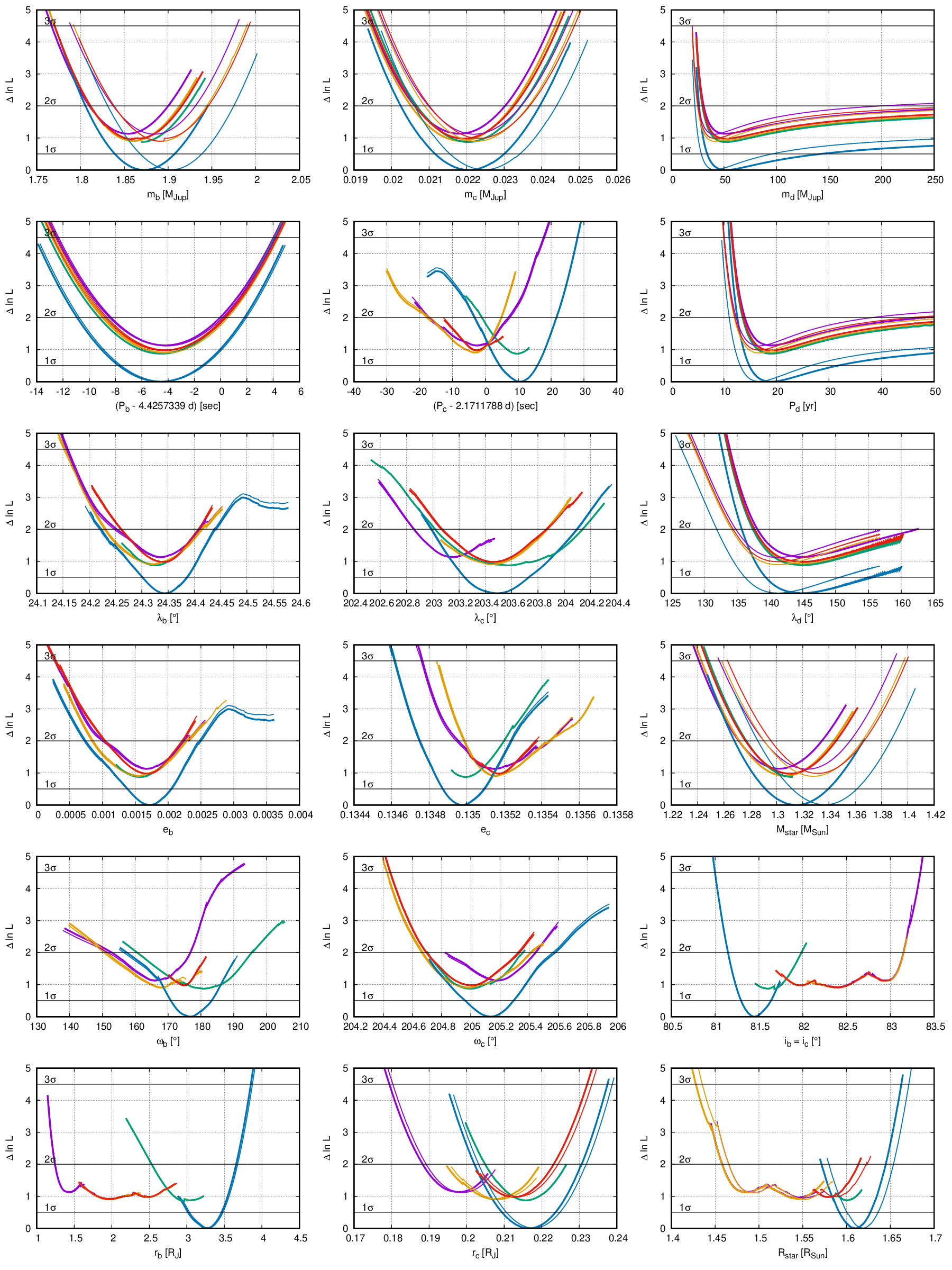}
\caption{Graphs of the log-likelihood ratio as a function of each of $18$ selected parameters of the TOI1408 three-planet model,
constructed for each of $9$ local extrema located (see text for a detailed explanation). Astrophysical estimations for the star
mass and radius are not taken into account here.}
\label{fig:pars2}
\end{figure}

We can confirm that orbital model given by \citet{Korth24}, or ones close to, can effectively explain the observed TTVs. However
our results appear more pessimistic regarding the parametric uncertainties. We found that the likelihood function has numerous
local maxima, basically demonstrating a ``rippled'' shape near its top. This is shown in Fig~\ref{fig:pars2} that demonstrates
so-called profile-likelihood curves for each of the $18$ system parameters: planetary masses $m_{b,c,d}$ (true ones, not $m\sin
i$), orbital periods $P_{b,c,d}$, mean longitudes $\lambda_{b,c,d}$, orbital eccentricities $e_{b,c}$, pericentra arguments
$\omega_{b,c}$, planet radii $r_{b,c}$, common orbital inclination $i_b=i_c$, star radius $R_\star$ and mass $M_\star$. Each
profile-likelihood curve shows the log of the likelihood ratio statistic $\Delta\log L$ as a function of a given parameter, i.e.
the log-likelihood maximized with respect to all the remaining free parameters (fixing the selected one), and subtracted from the
global maximum. In such a way all the maxima turn into minima inverting the graph.

In case if there was only a single extremum, the parametric uncertainty for a given parameter could be determined in the classic
way, that is by the abscissa range in which $\Delta \log L$ remains below some threshold determined by the asymptotic $\chi^2$
test. Namely, each $n\sigma$ threshold would require $\Delta \log L\leq n^2/2$, e.g. those thresholds should be $0.5$, $2.0$, and
$4.5$ for the $1\sigma$, $2\sigma$, and $3\sigma$ significance levels, respectively. Because of multiple extrema this rule will
need a correction which is considered below, but before that we will use such simplified thresholds as some rough reference
levels.

Different local extrema are best separated along the parameters $r_b$, $i_{b,c}$, and $R_\star$. Notice that these three
parameters jointly define the planet~\emph{b} transit geometry, so this phenomenon is likely related to the near-tangential
character of the transit. Simultaneously these parameters attain considerably larger uncertainties compared to \citet{Korth24}.
For example, the planet radius $r_b$ still appears rather poorly determined according to our analysis (from $\sim 1$~$R_J$ to
$\sim 4$~$R_J$). It looks as if the solution from \citet{Korth24} corresponded to just a single local maximum in the middle of
the range. Additionally, almost every such local extremum appears split into two ones differing by the parameters $M_\star$,
$P_d$ and $\lambda_d$. We reliably identified total of $9$ likelihood maxima (one of them did not to have a pair along the
$M_\star$ axis).

Notice that the multidimensional geometry of the likelihood function appears quite complicated. For example, the latter unpaired
extremum exists in quite a limited parametric range (green curves in Fig~\ref{fig:pars2}). Geometrically it would probably look
like a ``niche'' on the multi-dimensional surface of the likelihood function. Some other extrema may also disappear whenever some
critical parameter moves beyond certain limit; after that the optimization algorithm converges to another local extrema rendering
a jump in the corresponding plot. The part of the curve after such jump would simply duplicate another curve belonging to another
local extremum, so in Fig~\ref{fig:pars2} we cleaned such jumps away for the sake of clarity. Each curve stops being plotted
after a jump, if the jump exceeded a threshold of $0.1$.

One may notice that some extrema associate to apparently less realistic values of $R_\star$ and $M_\star$, so some of them may be
rejected by taking into account independent astrophysical estimations for these stellar parameters. \citet{Korth24}, based on
HARPSN spectra, estimated $M_\star=(1.31\pm 0.01)$~$M_\odot$ and $R_\star=(1.53\pm 0.02)$~$R_\odot$. This appears quite different
from e.g. \citet{Galazutdinov23} who relied on more noisy spectra and derived a smaller $R_\star$. However, in a recent
reanalysis of the HARPSN spectra \citet{Galazutdinov26} obtained $M_\star=(1.35\pm 0.04)$~$M_\odot$ and $R_\star=(1.50\pm
0.05)$~$R_\odot$. These values are in a good agreement with \citet{Korth24} but they have larger uncertainties, probably because
\citet{Galazutdinov26} tried to take into account possible model-related errors by joining multiple stellar evolution models.
Thanks to that we assume these newer estimations may be more realistic and use them in our analysis.

\begin{figure}
\centering
\includegraphics[width=\textwidth, angle=0]{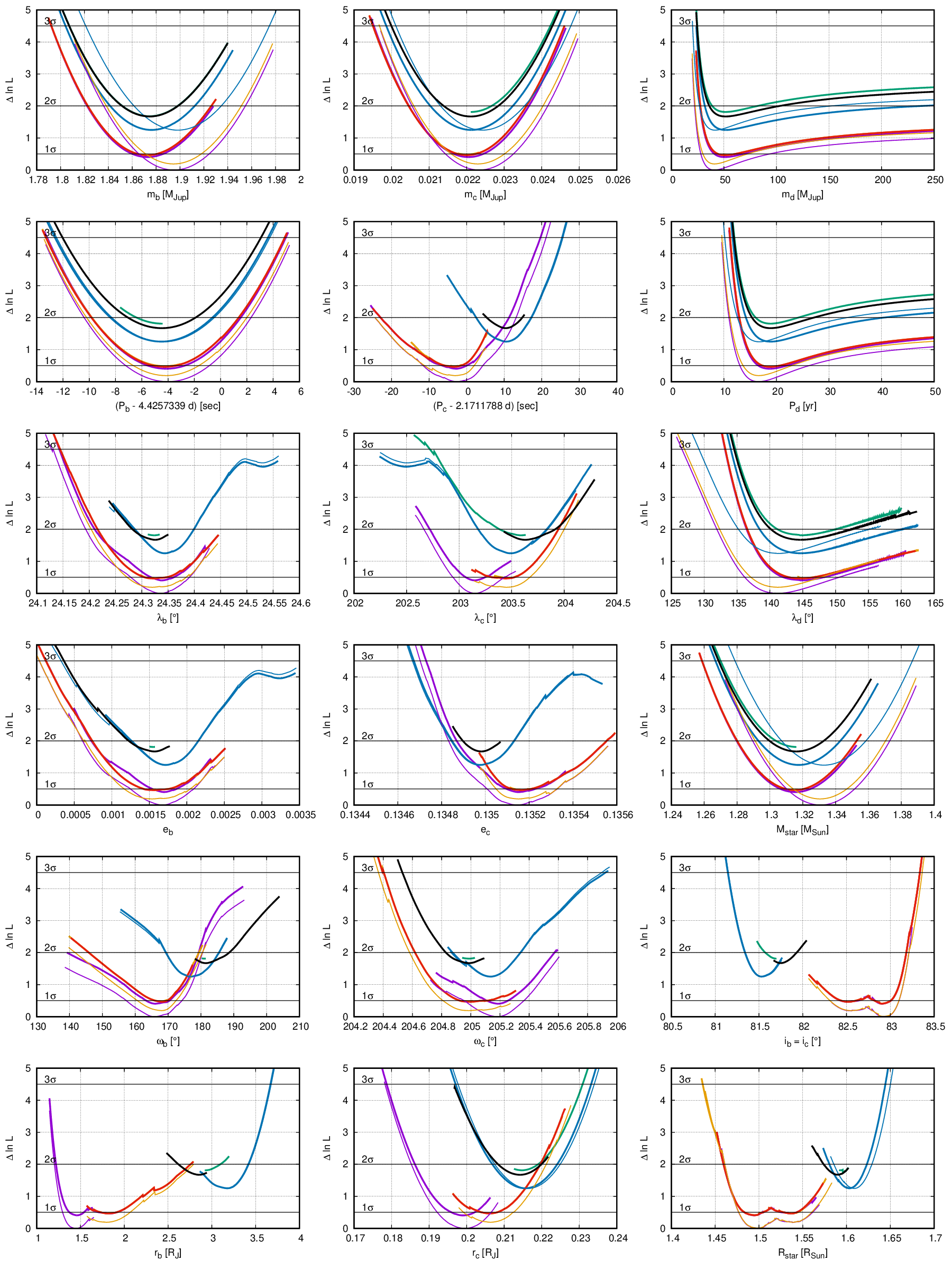}
\caption{Same as Fig~\ref{fig:pars2}, but taking into account astrophysical estimations for the star mass and radius (assuming no
correlation between them). See text for details.}
\label{fig:pars3}
\end{figure}

We considered the likelihood penalized by the 2D Gaussian distribution of $M_\star$ and $R_\star$. Such penalized likelihood is
proportional to the Bayesian posterior distribution density, assuming the prior for $M_\star$ and $R_\star$ as specified above.
Notice that neither \citet{Korth24} nor \citet{Galazutdinov26} provided a correlation between $M_\star$ and $R_\star$, so we
initially assumed them uncorrelated. In this case, after the penalizing, most of the likelihood extrema are preserved, although
one can see some changes in their position and shape (see Fig~\ref{fig:pars3}). In particular, larger $r_b$, as well as larger
$R_\star$ and smaller $i_{b,c}$, got moderately penalized, however one cannot say that the uncertainty ranges changed
drastically, so contrary to \citet{Korth24} planet~\emph{b} radius still remains rather poorly constrained.

\begin{figure}
\centering
\includegraphics[width=0.49\textwidth, angle=0]{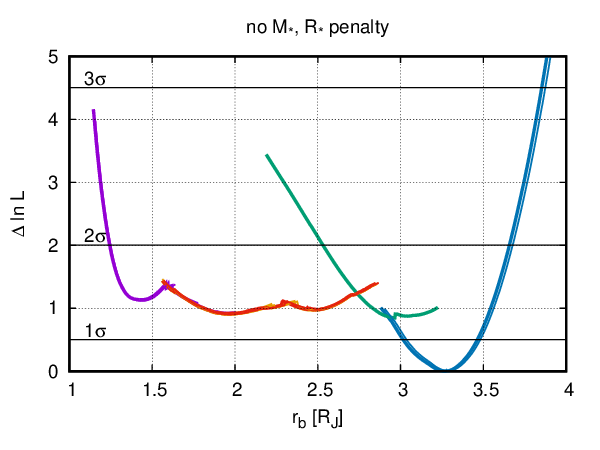}
\includegraphics[width=0.49\textwidth, angle=0]{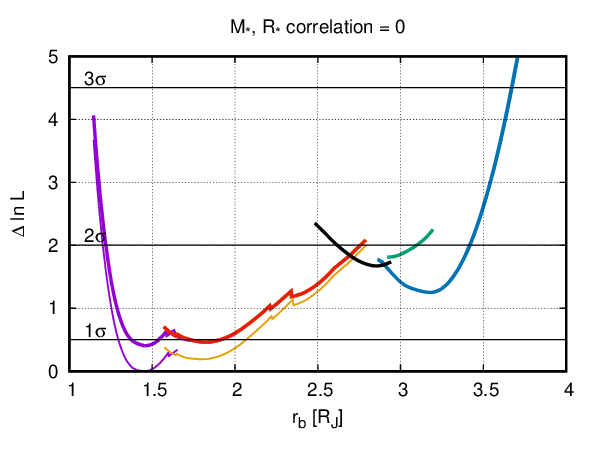}\\
\includegraphics[width=0.49\textwidth, angle=0]{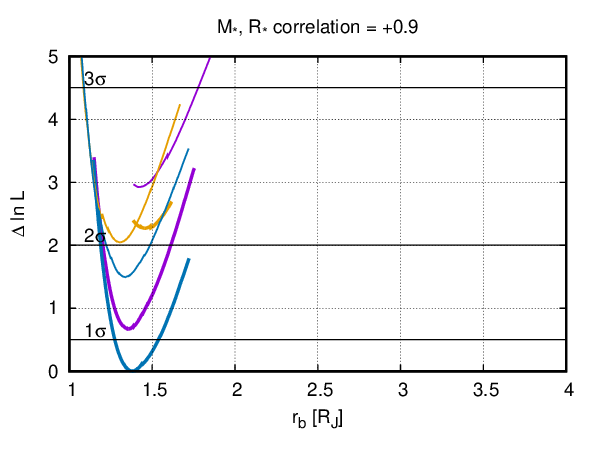}
\includegraphics[width=0.49\textwidth, angle=0]{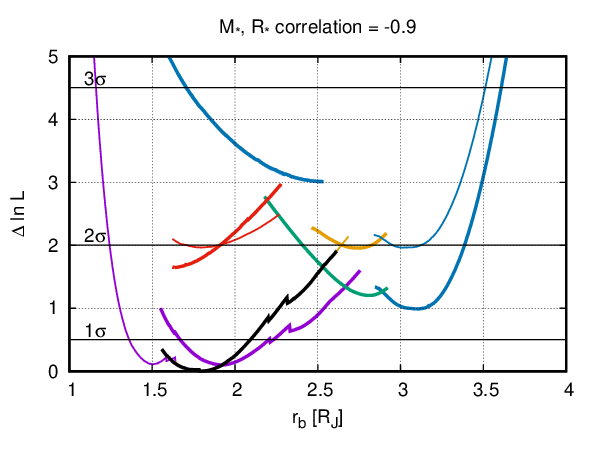}
\caption{The log-likelihood graphs for planet TOI-1408~\emph{b} radius assuming different correlation between the star mass and
radius. The top pair of plots duplicate the $r_b$ ones from Figs~\ref{fig:pars2} and~\ref{fig:pars3}, for a comparison.}
\label{fig:rb}
\end{figure}

After that we tried to take into account possible correlation between $M_\star$ and $R_\star$. We considered two additional test
cases, in which the correlation was set to either $-0.9$ or $+0.9$ (still assuming the 2D distribution to be Gaussian). For these
cases we demonstrate the likelihood-ratio graphs only for the planet radius $r_b$, see Fig~\ref{fig:rb} (which also replicates
the $r_b$ graphs from Figs~\ref{fig:pars2} and~\ref{fig:pars3}). The effect of this correlation on the likelihood is essential,
in particular the number of local extrema got roughly halved when this correlation was positive, but got increased when the
correlation was negative. This observation has two important consequence: (i) the orbital model for this system depends
significantly on the adopted prior for $M_\star$, $R_\star$, and (ii) we currently do not have a reliable enough prior for these
parameters. For example, one can see from Fig~\ref{fig:rb} that the uncertainty in $r_b$ may reduce whenever we assume a large
positive correlation between $M_\star$ and $R_\star$, however currently it does not seem that we have a solid ground to make such
an assumption.\footnote{Notice that \citet{Galazutdinov23} reported the correlation between their estimations for $M_\star$ and
$R_\star$ about $+0.87$ but this appears unreliable, because correlations between spectrum parameters $T_{\rm eff}$, $\log g$,
[Fe/H] were not taken into account.} Another point here is that the effect of the unknown $M_\star$ and $R_\star$ correlation on
the model is comparable to the effect of the entire $(M_\star, R_\star)$ penalty itself.

As one can see, the effect of the astrophysical priors of $M_\star$ and $R_\star$ may be important, for example it can shift
different local maxima vertically, and may affect their total number. But on the other hand this effect appears paradoxically
unpredictable because important information is not available. In addition to the lacking correlation between $M_\star$ and
$R_\star$, our initial assumption that the 2D distribution of $(M_\star,R_\star)$ is Gaussian may be (and likely is) inaccurate.
In other words, while our astrophysics-derived knowledge about the $M_\star$ and $R_\star$ is significant, it still cannot be
formalized at an accuracy level required by the model fitting task.

\subsection{Three-planet orbital fits}
\label{sect:fits}

From our analysis above it follows that the likelihood-ratio graphs discussed above (Figs~\ref{fig:pars2}--\ref{fig:rb}) cannot
be directly used to derive parametric uncertainties, mainly because of the multiple extrema effect. The 1,2,3-$\sigma$ threshold
levels were shown there just for a rough reference, but they were constructed assuming an asymptotic $\chi^2$ distribution for
$2\Delta\log L$ (by the Wilks theorem), which would only hold for a linearizable task (whenever there is a single local extremum
of $\Delta\log L$ having a near-parabolic shape). Therefore, to obtain more realistic parametric uncertainties these significance
levels need to be corrected.

First of all let us briefly formalize the issue in the classic (single-extremum) treatment. The log-likelihood-ratio is defined
as
\begin{equation}
Z(p_{\rm test}) = \Delta \log \mathcal L = \max_{p_{\rm test}, \mathbf p_{\rm other}} \log \mathcal L(p_{\rm test},\mathbf p_{\rm other}) - \max_{\mathbf p_{\rm other}} \log \mathcal L(p_{\rm test},\mathbf p_{\rm other}) \geq 0,
\label{llr}
\end{equation}
where $\mathcal L$ is the likelihood, $p_{\rm test}$ is a test parameter under study, while $\mathbf p_{\rm other}$ are all the
other parameters being optimized out. Then the maximum-likelihood estimation of $p_{\rm test}$ is
\begin{equation}
\hat p_{\rm test} = \arg \min Z(p_{\rm test}),
\end{equation}
while the uncertainty ranges for $p_{\rm test}$ can be approximated by the set:
\begin{equation}
C(\alpha) = \{ p_{\rm test}: Z(p_{\rm test})\leq z_{\rm thr}(\alpha) \},
\label{conf}
\end{equation}
where $z_{\rm thr}(\alpha)$ depends on the desired confidence probability $\alpha$ (complementary p-value) in such a way that
$C(\alpha)$, which is a random domain, covers the true value of $p_{\rm test}$, $p_{\rm test}^*$, with this probability $\alpha$:
\begin{equation}
\Pr\{ C \ni p_{\rm test}^*\} = \alpha.
\label{csig}
\end{equation}
Moreover, from the Wilks theorem it follows that $2z_{\rm thr}(\alpha)$ should be asymptotically (for $N\to\infty$) an
$\alpha$-quantile of a $\chi^2$-distribution with the number degrees of freedom being equal to $\dim p_{\rm test}=1$. Therefore,
this should be the $\chi^2(1)$ distribution, implying that $z_{\rm thr}$ attains a simple relationship with quantiles of the
standard normal distribution, namely $z_{\rm thr}=s^2/2$, where $s$ is that quantile. In this layout the uncertainty intervals
corresponding to e.g. $1\sigma$, $2\sigma$, and $3\sigma$ significance (offering $1-\alpha=0.32$, $0.046$, and $0.0027$,
respectively) are defined by setting $z_{\rm thr}=0.5$, $2.0$, and $4.5$, respectively. These exact levels were shown in
Figs~\ref{fig:pars2}--\ref{fig:rb} for a reference.

But in our case the maxima involved in~(\ref{llr}) become multivalued, so $Z$ becomes multivalued as well. In fact, we have a set
of multiple overlapping branches $Z_k(p_{\rm test})$ which can be defined like~(\ref{llr}) but restricting each maximization to a
$k$th local maximum.\footnote{Notice that two maximizations in~(\ref{llr}), with and without maximizing over $p_{\rm test}$,
should both refer to the same branch then. This ``branch'' can be formally defined in the sense of a smooth change of the
parameters along it.} In Figs~\ref{fig:pars2}--\ref{fig:rb} we plotted these $Z_k$, though shifting them vertically as a whole in
such a way that the global maximum is set at zero ordinate. Since we have multiple $Z_k$ branches now, all the entities
introduced above become branch-specific as well, in particular we have individual estimations $\hat p_{\rm test}^{(k)}$ and
individual ranges $C_k$ per each $Z_k$.

Based on the study from Sect.~\ref{sect:3pmod}, practically none of the local maxima in the initial (unpenalized) likelihood can
be rejected. But simultaneously in Figs~\ref{fig:pars2}--\ref{fig:rb} one may notice that individual $Z_k$ often preserve a
near-parabolic shape. This indicates that the task still remains locally linearizable, so the classic theory may still be
applicable to individual $Z_k$ in some sense. However, different such branches behave like independent alternative models that
are fitted simultaneously, so the significance threshold $z_{\rm thr}(\alpha)$ should be corrected to take into account the
effect of multiple testing. By using such an approach we consider linear and nonlinear effects at different levels of the
likelihood function topology.

To treat the effect of maxima multiplicity we follow a conservative approach, assuming that any of $Z_k$ branches may cover the
true parameter value $p_{\rm test}^*$. So all branches are treated equally and if at least a single condition in the left hand
side of~(\ref{csig}), for any $Z_k$, is violated we treat this as a ``miss case'' even if all the other similar conditions are
OK. This formally means that the left hand side of~(\ref{csig}) should be replaced by the probability of a
\emph{conjunction} of all similar branch-specific events. Our additional simplifying assumption will be that all these events are
statistically independent (meaning that branch-specific estimations are independent as well). Then~(\ref{csig}) turns into a
product of probabilities:
\begin{equation}
\alpha = \Pr\left\{ \bigwedge_{k=1}^{N_{\rm branches}} C_k \ni p_{\rm test}^* \right\} \simeq \alpha_1^{N_{\rm branches}}, \quad \alpha_1 = \Pr\{ C_k \ni p_{\rm test}^* \}.
\end{equation}
This implies that to achieve the desired $\alpha$ level in the \emph{multi-extrema} treatment we should aim to have the classic
(single-extremum) probability of $\alpha_1=\alpha^{1/N_{\rm branches}}$. The confidence set itself is then the union of all
$C_k(\alpha_1)$.

In fact this approach is a simplified adaptation of the correction known as the bandwidth penalty in periodogram analysis
\citep{HorneBal86,Koen90,Baluev08a,Baluev13b}, and in a more general setting it is known as the multiple hypothesis testing or
the look-elsewhere effect \citep{Bayer21}. The number of branches $N_{\rm branches}$ is a bit tricky element of this approach. To
be more strict, it should be understood in the average sense, but it can also be estimated based on the actual number of local
maxima observed in the likelihood of a particular task. In the analysis above we observed $N_{\rm branches}\sim 10$, but this
number may need some correction. On the one hand, there is still a risk that we missed some extrema and thus undervalued it, but
on the other hand the branches that we identified are unlikely to be all independent. Moreover, some of them are only partly
separated from the others and some span rather narrow parametric ranges. Since our approach is already conservative by its
design, in order to avoid being overconservative we adopted $N_{\rm branches}\sim 5$. Then, for example, to achieve the
post-correction $\alpha$ probabilities of $68\%$, matching the $1\sigma$ level, one should require $\alpha_1$ to be such that it
maps to the (uncorrected) $1.8\sigma$ level.

Notice that effective $N_{\rm branches}$ is a somewhat subjective choice here; this particular value of $5$ was partly motivated
by the observation that for the parameters $r_b$, $i_{b,c}$, and $R_\star$ most of the branches are grouped into close pairs, so
the number of independent branches should be about half of the total number. The effect of $N_{\rm branches}$ on the resulting
ranges is nonlinear and does not appear very large, for example assuming $N_{\rm branches}=10$ the corrected $1\sigma$ range maps
to the uncorrected $2.1\sigma$; for many parameters this results in only a rather small range change because of a high slope of
the likelihood graph.

\begin{table}
\bc
\begin{minipage}[]{100mm}
\caption[]{Parameters of the three-planet model for TOI-1408 (reference epoch
$T_0=2459987.815534$~${\rm BJD}_{\rm TDB}$, Jacobi coordinates)\label{tab:pars}}\end{minipage}
\setlength{\tabcolsep}{2pt}
\small
\begin{tabular}{|c|c|c|c|}
\hline
parameter   & planet~\emph{b}       & planet~\emph{c}   & companion~\emph{d} \\
\hline
$m$, $M_{\rm Jup}$  & $   1.881    \left( +0.070   \atop  -0.060   \right) $ & $   0.0222  \left(+0.0019    \atop -0.0017 \right) $ & $  48 \left(+\infty \atop   -25\right) $ \\
$r$, $R_{\rm Jup}$  & $   2.3      \left( +1.2     \atop  -1.1     \right) $ & $   0.209   \left(+0.020     \atop -0.023  \right) $ & \\
$P$, d              & $   4.425684 \left( +0.000069\atop  -0.000069\right) $ & $   2.17121 \left(+0.00021   \atop -0.00031\right) $ & $6600 \left(+\infty \atop -2500\right) $ \\
$\lambda$, $^\circ$ & $  24.33     \left( +0.11    \atop  -0.12    \right) $ & $ 203.43    \left(+0.81      \atop -0.71   \right) $ & $ 144 \left(+\infty \atop   -11\right) $ \\
$e$                 & $   0.00162  \left( +0.00090 \atop  -0.00093 \right) $ & $   0.13508 \left(+0.00049   \atop -0.00031\right) $ & \\
$\omega$, $^\circ$  & $ 172        \left(+27       \atop -34       \right) $ & $ 205.06    \left(+0.51      \atop -0.46   \right) $ & \\
\cline{2-4}
$i$, $^\circ$       & \multicolumn{3}{c|}{$      82.21  \left(+0.99  \atop -0.95  \right) $} \\
$M_\star$, $M_\odot$& \multicolumn{3}{c|}{$       1.321 \left(+0.049 \atop -0.042 \right) $} \\
$R_\star$, $R_\odot$& \multicolumn{3}{c|}{$       1.556 \left(+0.082 \atop -0.098 \right) $} \\
\hline
\end{tabular}
\ec
\end{table}

The results of this computation are shown in Table~\ref{tab:pars}. The nominal values of parameters were obtained plainly by
averaging between all the local extrema, while the uncertainty ranges were obtained using the method described above. Notice that
we did not use the astrophysical constraints on $M_\star$, $R_\star$ here, in particular this table provides a purely dynamical
estimation for the star mass. This estimation of $M_\star$ appears quite similar to the astrophysical one by
\citet{Galazutdinov26}.

\begin{figure}
\centering
\includegraphics[width=0.49\textwidth, angle=0]{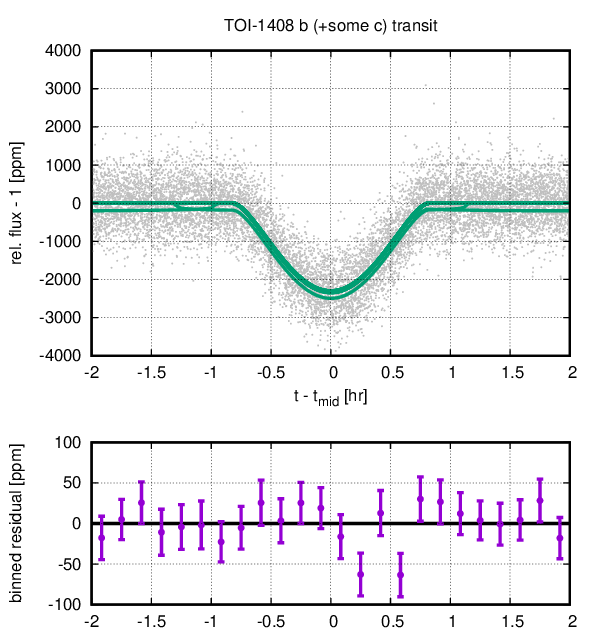}
\includegraphics[width=0.49\textwidth, angle=0]{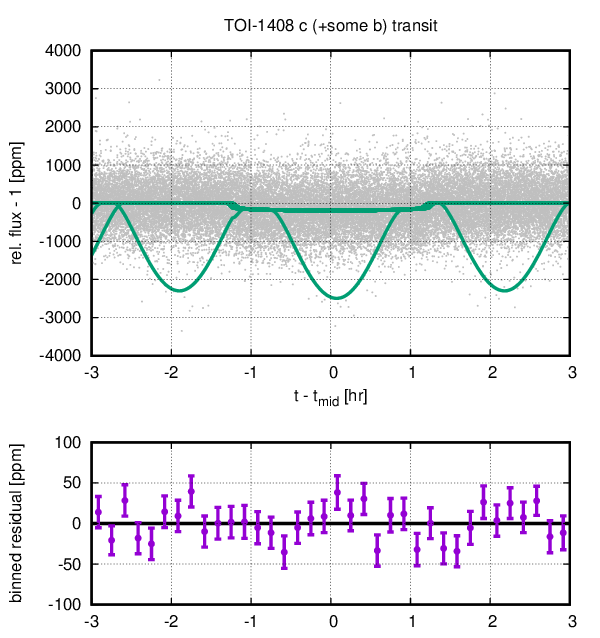}
\caption{In top panels transits of TOI1408~\emph{b} (left) and~\emph{c} (right) are shown along with their best fitting models
(computed for one of likelihood maxima). In panels below the corresponding binned residuals and their uncertainties are plotted.
Notice that there are a few occurrences when planets~\emph{b} and~\emph{c} transits partly overlapped with each other; in such a
case all transit curves are plotted in the same graph.}
\label{fig:transits}
\end{figure}

In Fig~\ref{fig:transits} we show TESS data photometric residuals computed for one of the likelihood local maximum. These plots
appeared visually identical for all the maxima, so we show just one of them. The data were phased to the transit midtimes of
either planet~\emph{b} or planet~\emph{c}, and the polynomial trends were subtracted to improve the graphs clarity. Notice that
planet \emph{c} transit phase slowly circulates relative to the planet~\emph{b} one, therefore sometimes we meet a situation when
both transits occur simultaneously. In case of such an occurrence we did not subtract the contribution of the other transit in
Fig~\ref{fig:transits} to render this effect itself.

Notice that binned residuals, also shown in Fig~\ref{fig:transits}, do not apparently reveal any remaining variation within the
duration of either transit, their behaviour inside and outside of a transit looks qualitatively identical. This indicates that
our models do not leave any visible systematic errors. The uncertainties of the binned residuals near the transits of
planet~\emph{b} appear systematically larger than those near the transits of planet~\emph{c} because of the different number of
lightcurves (and data) stacked in the corresponding plots. This does not reflect any phase-dependency effect.

\subsection{Discussion of subtle effects that were neglected}

\subsubsection{Orbital non-coplanarity}

In the above analysis we considered only coplanar models of the system. Non-coplanar configurations were considered by
\citet{Korth24} by allowing different inclinations $i_b$ and $i_c$. However, their inclination estimations appeared the same
within the uncertainties. Simultaneously, their model assumed strictly coinciding orbital nodes which means that it was not
entirely noncoplanar. We tried to fit entirely noncoplanar configurations with an additional free angle for the inter-orbital
ascending node. This angle appeared rather robustly fittable, and it always was consistent with zero (below $1^\circ$), as well
as the mutual inclination between planets~\emph{b} and~\emph{c} orbits. However, the issue of multiple likelihood extrema was
significantly exacerbated, requiring a larger amount of numeric computations. Because of that we did not consider non-coplanar
configurations in this study in detail. Currently we did not detect any hints of an interorbital inclination, but this issue
perhaps needs additional investigation in future.

\subsubsection{R\o{}mer delay because of the outermost body}

A distant orbiting body may cause a time delay of the observed light arrival because of a combination of a finite light speed
with the star motion about the system barycenter. This effect was initially observed in the Solar System and is often named after
Ole R\o{}mer. Such an effect may be important in some extrasolar systems where it can cause an apparent TTV \citep{Harre23}. In
case of TOI-1408 this effect may come from the outermost body \emph{d}.

Notice that to predict the time delay variation it is unnecessary to accurately know the outer body orbit and mass. One can just
use the observed non-linear Doppler trend which is constrained quite well. The R\o{}mer effect can be determined by integrating
this Doppler trend, which yields the star radial coordinate, and then dividing this integral by the speed of light.

\begin{figure}
\centering
\includegraphics[width=0.8\textwidth, angle=0]{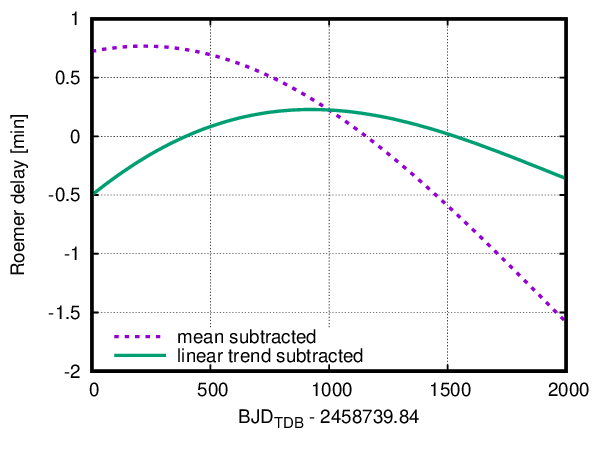}
\caption{The R\o{}mer effect caused by the outermost body ``\emph{d}'' on the inner subsystem of two planets.}
\label{fig:td}
\end{figure}

The resulting variation is shown in Fig~\ref{fig:td}. Notice that this variation has a free offset (the integration constant)
tied to the total time the light needs to reach the observer. Also, this variation involves a freely variable linear trend
because of a relative nature of Doppler data. The latter effect depends on the absolute radial velocity and simply scales all the
orbital periods up by the same near-unit coefficient (or, alternatively, scales the time unit). In practice such a correction is
rarely applied, even if it exceeds the periods uncertainties. Both these linear effects are not important in our study, so to
keep only the essential non-linear part of the time delay variation we subtracted the best fitting linear trend from it. In
Fig~\ref{fig:td} one can see that after such a subtraction the effect remained below $0.5$~min over the entire observation range.
This is well below the typical uncertainty of a single TESS transit timing ($\gtrsim 2$~min) and should not affect our analysis.
Nevertheless, to additionally verify that we applied our time correction from Fig~\ref{fig:td} to all input data and rerun all
the fits. It appeared that all the parameters shifted only negligibly, by a quantity no more than $\sim 0.1$ times the respective
parameter uncertainty (even before their correction for multiple extremes). Therefore, presently the R\o{}mer effect does not
seem to be significant in TOI-1408 system, although it can appear significant in future when the observation time base is
expanded.

\subsubsection{Rossiter-McLaughlin effect}

\citet{Korth24} and \citet{Galazutdinov26} both gave $v\sin i \sim 10$~km/s based on the spectrum models, and considering this
value together with the derived planet~\emph{b} transit depth and the limb-darkening near the stellar limb, the expected maximum
amplitude of the Rossiter-McLaughlin (RM) effect should be $\sim 10$~m/s. This is comparable to the uncertainty of a single
HARPSN measurement (summing up all its modelled noises). Therefore, the effect would perhaps be detectable had it that large
magnitude. However, given the near-grazing transit geometry, the maximum magnitude is reached only for specific misalignment
angle $\lambda=90^\circ$. If the planet~\emph{b} orbit and stellar equator are aligned then the planet would transit only the
near-pole region of the star, resulting in a negligibly small RM effect. Regarding the planet~\emph{c}, its RM effect should be
$\sim 1$~m/s at most, which is negligible regardless of the orbital alignment.

Our attempt to estimate the RM effect from the Doppler data (including both planets) implied, formally, $v\sin i \sim 15$~km/s,
with the uncertainty comparable to the value, therefore this effect did not appear reliably detectable. Contrary to that, the
formal estimation for the alignment angle $\lambda$ appeared apparently informative, namely $0^\circ \pm 7^\circ$ (rather
independent on the likelihood extremum being chosen). This might be a null-detection constraint actually, that is $\lambda$ is
estimated to be small because otherwise the RM effect magnitude would be large which is not the case. Nevertheless the
planet~\emph{b} orbit appears more likely to be aligned with the star equator, however this requires an additional detailed
analysis of several accompanying issues (e.g. of the ``spectral'' limb darkening law). Since the RM effect was not a primary goal
of our study, we did not investigate this further. However, we ensured that adding or removing it to our models does not
significantly change the best fitting parameters. The maximum shift appeared just $\sim 0.05$ times the corresponding parameter
uncertainty (even before their correction for multiple extrema). This included the effect from both transiting planets,
TOI-1408~\emph{b} and~\emph{c}.

\subsubsection{Uncertainties of the limb darkening law}

Finally, we considered the effect of the LDL uncertainties on the fit parameters. We perturbed each of quadratic LDL coefficients
by $\pm 1\sigma$, assuming the level-1 uncertainties from Table~\ref{tab:ldc}. In this case the maximum relative shift of the
best fitting parameters was only $0.06$ times the given parameter uncertainty (again before their correction for multiple
extrema). This is entirely negligible. Though the LDL has additional systematic errors which are more difficult to characterize
accurately, it does not seem likely that their effect may grow so large to become significant for our analysis.

\section{Summary and discussion}
\label{sect:disc}

In this work we performed a detailed reassessment of the TOI-1408 planetary system architecture, including additional new TESS
data as well as involving a more detailed analysis of the limb darkening, photometric and Doppler noises, and the structure of
the derived likelihood function. We generally confirm the three-body system interpretation by \citet{Korth24}, with two inner
planets being at least close to the 2:1 mean-motion resonance and causing TTV effects on each other. However we derived
considerably larger parametric uncertainties, in particular the planet~\emph{b} radius still remains ill-determined according to
our treatment. The reason for such difference with \citet{Korth24}, as we see it, is probably because they adopted rather
narrowly localized priors for several parameters. Not only the star mass and radius were assumed to have a narrow prior, but a
few other parameters as well, and the motivation for that is not always obvious.

Notice that for a two-planet dynamics to be termed resonant it is usually assumed enough if at least one critical angle librates
\citep{Malhotra-Resonances}. For all our local likelihood maxima we verified that this remains nearly the same as in
\citet{Korth24}, that is one critical angle (for the planet~\emph{c}) librates with the period $\sim 138$~d, while the other one
(for the planet~\emph{b}) circulates with the period of $\sim 295$~d. Clearly, the primary $138$~d TTV period seen in
Fig~\ref{fig:ttv} for both the planets is owed to the libration, while the circulation of the second angle is not seen in TTV but
it results in a drift of planetary transit phases relative to each other (so sometimes two transits overlap, sometimes not).
However \citet{Korth24} performed a more detailed dynamical analysis and did not find the resonance separatrix in the
Hamiltonian, implying that the motion is not resonant in a more strict dynamical sense. In the context of this work this
difference between two definitions of a resonance is not very important, as considering the dynamics of TOI-1408 in detail was
not among our goals.

Moreover, to certainly resolve the issue of whether the motion is resonant or not, it is necessary to consider the effect of
parametric uncertainties on the system dynamics. This question was not considered here, as well as by \citet{Korth24}, at least
not comprehensively. Nevertheless, any reasonable orbital solution has to keep the robustly observable TTV signal from
Fig~\ref{fig:ttv}. It is possible, in principle, that entirely non-resonant configurations (with both critical angles
circulating) exist, simultaneously explaining the observed TTV. However, such a qualitative transition from librating dynamical
regime to a circulation-only one would not appear smooth enough in terms of the likelihood function. Possible circulating
solutions, if they exist, would necessarily appear as separate likelihood modes, but modes that we found all correspond to
librating solutions. Hence we remain sceptical regarding the possibility that this TTV signal could be reproduced without
preserving the libration dynamics of the same qualitative type.

However, the effect of parametric uncertainty on the existence of a resonance separatrix is less obvious. The separatrix may, in
theory, appear if we step some away from the nominal solution, and simultaneously the goodness of the fit may be possibly kept
reasonable as long as such a change does not trigger abrupt change of the dynamical regime on the observation time range. This
question needs a further study.

The parameters of the third body TOI-1408~\emph{d} that manifests itself through a nonlinear Doppler trend still cannot be
constrained well even in the simplified approximation of a circular orbit. Its orbital period appears indistinguishable from the
infinity, as well as this object mass. However, lower limits can be placed on these parameters that cast doubts on its planetary
nature. Nevertheless it is still possible that either (i) the system contains multiple long-period objects responsible for the
observed cumulative Doppler trend, and in such a case some or even all of them still can have planetary masses, or (ii) the
companion's orbit may be eccentric which violates our initial assumption. By admitting nonzero $e_d$ the lower limit on $m_d$ may
reduce, however such a model is difficult to process in our task because it would have a high level of degeneracy.

Regarding the astrophysical star mass and radius estimations for TOI-1408, we believe their uncertainties given by
\citet{Korth24} may be overly optimistic because they did not consider systematic errors owed to differences between stellar
models (their systematic errors). In this work we used the star mass and radius estimations from \citet{Galazutdinov26} who
considered multiple stellar models and obtained significantly larger uncertainties for $M_\star$, $R_\star$. However, another
issue appeared here because the prior distribution for $M_\star$, $R_\star$ has to be characterized in considerable detail to be
useful with TOI-1408. For example the planetary model depends significantly on the correlation between $M_\star$ and $R_\star$,
which is usually not provided. Here the underlying issue is because the analysis involves a set of nonlinear models chained to
each other sequentially, namely raw spectra are first fitted to obtain $T_{\rm eff}$, $\log g$, and [Fe/H], then these stellar
parameters are used as input data for stellar evolution models to obtain $M_\star$ and $R_\star$, and finally these $M_\star$ and
$R_\star$ are used as input data for our three-planet model fit. However, in each link of this chain one has to make simplifying
assumptions regarding all the intermediary parameters, in particular their correlations are omitted and their distribution is
assumed to be Gaussian. In the case of TOI-1408 a better and more reliable way would probably be to perform a self-consistent fit
of a combined model involving all the underlying models simultaneously, namely the transit and Doppler models, as well as stellar
evolution models and raw spectrum models, thus using the raw spectra as input data. In this way one would avoid intermediary
estimations, as well as any assumptions about them.

We also performed a detailed investigation of the TOI-1408 limb darkening and constructed a corrected quadratic model tuned to fit
better the near-limb zone, which is transited by the planet~\emph{b}. We provided these models for a set of $18$ passbands,
including the TESS one as well as classic photometric systems typically used in ground-based observations. Based on this
investigation one may argue that possible remaining LDL errors are likely negligible in the context of modelling TOI-1408
exoplanetary transit curves.

Regarding the noises seen in Doppler data, we did not reveal clear hints of stellar rotation period in them. It seems that in the
most cases they are more likely to have an instrumental nature, while for the best-accuracy HARPSN dataset there is only very
modest correlated noise with no convincing signs of rotational modulation.

\normalem
\begin{acknowledgements}
We would like to thank the anonymous reviewer for their valuable suggestions that helped to improve the manuscript.
\end{acknowledgements}

\bibliographystyle{raa}
\bibliography{TOI1408}

\begin{thebibliography}{31}
\providecommand\natexlab[1]{#1}
\providecommand\JournalTitle[1]{#1}

\bibitem[Abubekerov \& Gostev(2013)]{AbubGost13}
Abubekerov, M.~K., \& Gostev, N.~Y. 2013, \mnras, 432, 2216

\bibitem[Baluev(2008)]{Baluev08a}
Baluev, R.~V. 2008, \mnras, 385, 1279

\bibitem[Baluev(2011)]{Baluev11}
Baluev, R.~V. 2011, Celest. Mech. Dyn. Astron., 111, 235

\bibitem[Baluev(2013{\natexlab{a}})]{Baluev13b}
Baluev, R.~V. 2013{\natexlab{a}}, \mnras, 431, 1167

\bibitem[Baluev(2013{\natexlab{b}})]{Baluev13a}
Baluev, R.~V. 2013{\natexlab{b}}, \mnras, 429, 2052

\bibitem[Baluev(2013{\natexlab{c}})]{Baluev13c}
Baluev, R.~V. 2013{\natexlab{c}}, \ac, 2, 18

\bibitem[Baluev(2015)]{Baluev14a}
Baluev, R.~V. 2015, \mnras, 446, 1493

\bibitem[Baluev(2018)]{Baluev18c}
Baluev, R.~V. 2018, \ac, 25, 221

\bibitem[Baluev \& Shaidulin(2015)]{BaluevShaidulin15}
Baluev, R.~V., \& Shaidulin, V.~S. 2015, \mnras, 454, 4379

\bibitem[Baluev \& Sokov(2025)]{BaluevSokov25}
Baluev, R.~V., \& Sokov, E.~N. 2025, \mnras, 537, 171

\bibitem[Baluev {et~al.}(2015)]{Baluev15a}
Baluev, R.~V., Sokov, E.~N., Shaidulin, V.~S., {et~al.} 2015, \mnras, 450, 3101

\bibitem[Bayer {et~al.}(2021)]{Bayer21}
Bayer, A.~E., Seljak, U., \& Robnik, J. 2021, \mnras, 508, 1346

\bibitem[Bruntt {et~al.}(2010)]{Bruntt10}
Bruntt, H., Bedding, T.~R., Quirion, P.-O., {et~al.} 2010, \mnras, 405, 1907

\bibitem[Claret(2017)]{Claret17}
Claret, A. 2017, \aap, 600, A30

\bibitem[Claret \& Bloemen(2011)]{ClaretBloemen11}
Claret, A., \& Bloemen, S. 2011, \aap, 529, A75

\bibitem[Ferraz-Mello(1981)]{FerrazMello81}
Ferraz-Mello, S. 1981, \aj, 86, 619

\bibitem[Foreman-Mackey {et~al.}(2017)]{Foreman-Mackey17}
Foreman-Mackey, D., Agol, E., Ambikasaran, S., \& Angus, R. 2017, \aj, 154, 220

\bibitem[Galazutdinov {et~al.}(2026)]{Galazutdinov26}
Galazutdinov, G.~A., Korotin, S.~A., Poklad, D.~B., Kazanina, E.~I., \&
  Valyavin, G.~G. 2026, \mnras, submitted

\bibitem[Galazutdinov {et~al.}(2023)]{Galazutdinov23}
Galazutdinov, G., Baluev, R., Valyavin, G., {et~al.} 2023, \mnras, 525, L111

\bibitem[Harre \& Smith(2023)]{Harre23}
Harre, J.-V., \& Smith, A. M.~S. 2023, Universe, 9, id506

\bibitem[Horne \& Baliunas(1986)]{HorneBal86}
Horne, J.~H., \& Baliunas, S.~L. 1986, \apj, 302, 757

\bibitem[Koen(1990)]{Koen90}
Koen, C. 1990, \apj, 348, 700

\bibitem[Korth {et~al.}(2024)]{Korth24}
Korth, J., Chaturvedi, P., Parviainen, H., {et~al.} 2024, \apjl, 971, L28

\bibitem[Kov\'{a}cs {et~al.}(2002)]{Kovacs02}
Kov\'{a}cs, G., Zucker, S., \& Mazeh, T. 2002, \aap, 391, 369

\bibitem[Malhotra(2012)]{Malhotra-Resonances}
Malhotra, R. 2012, in Encyclopedia of Life Support Systems (EOLSS), ed.
  A.~Celletti (Paris, France: Eolss Publishers), developed under the auspices
  of the UNESCO

\bibitem[Mandel \& Agol(2002)]{MandelAgol02}
Mandel, K., \& Agol, E. 2002, \apj, 580, L171

\bibitem[Rajpaul {et~al.}(2015)]{Rajpaul15}
Rajpaul, V., Aigrain, S., Osborne, M.~A., Reece, S., \& Roberts, S. 2015,
  \mnras, 452, 2269

\bibitem[Southworth(2015)]{Southworth15}
Southworth, J. 2015, Astrophysics Source Code Library, ascl:1511.016

\bibitem[Valyavin {et~al.}(2022{\natexlab{a}})]{Valyavin22a}
Valyavin, G., Beskin, G., Valeev, A., {et~al.} 2022{\natexlab{a}}, Photonics,
  9, 950

\bibitem[Valyavin {et~al.}(2022{\natexlab{b}})]{Valyavin22b}
Valyavin, G., Beskin, G., Valeev, A., {et~al.} 2022{\natexlab{b}},
  Astrophysical Bulletin, 77, 495

\bibitem[Zechmeister \& K{\"u}rster(2009)]{ZechKur09}
Zechmeister, M., \& K{\"u}rster, M. 2009, \aap, 496, 577

\end{thebibliography}

\end{document}